\documentclass[letterpaper, 10 pt, conference]{ieeeconf}

\IEEEoverridecommandlockouts                              

\usepackage{cite}
\usepackage[T1]{fontenc}
\usepackage{amsmath,amssymb,amsfonts}
\usepackage{graphicx}
\usepackage{bm} 
\usepackage{textcomp}
\usepackage{amsmath}  
\usepackage{tikz}
\usepackage{xcolor}                                  
\usepackage{amsmath}
\usepackage{varioref}
\usepackage{hyperref}      
\usepackage{color}     
\usepackage{amssymb}
\usepackage{mathtools}
\usepackage{overpic}
\usepackage{subcaption}

\newtheorem{Hypothesis}{Hypothesis}
\usepackage{courier}
\usepackage{algorithm}
\usepackage{algpseudocode}
\usepackage{bm}
\def\BibTeX{{\rm B\kern-.05em{\sc i\kern-.025em b}\kern-.08em
    T\kern-.1667em\lower.7ex\hbox{E}\kern-.125emX}}
\begin{document}

\title{\LARGE \bf Early Versus Late Traffic Management For Autonomous Agents\\

\author{Salman Ghori, Ania Adil, Eric Feron
\thanks{Salman Ghori, Ania Adil, and Eric Feron are with Computer, Electrical and Mathematical  Science $\&$ Engineering Division (CEMSE), KAUST, Saudi Arabia. (email:{\tt\footnotesize salman.ghori@kaust.edu.sa, ania.adil@kaust.edu.sa, eric.feron@kaust.edu.sa} }}

}
\maketitle

\begin{abstract}
Intersections pose critical challenges in traffic management, where maintaining operational constraints and ensuring safety are essential for efficient flow. This paper investigates the effect of intervention timing in management strategies on maintaining operational constraints at intersections while ensuring safe separation distance, avoiding collisions, and minimizing delay. We introduce control regions, represented as circles around the intersection, which refers to the timing of interventions by a centralized control system when agents approach the intersection. We use a mixed-integer linear programming (MILP) approach to optimize the system's performance. To analyze the effectiveness of early and late control measures, a simulation study is conducted, focusing on the safe, efficient, and robust management of agent movement within the control regions.

\end{abstract}


\section{Introduction}

Autonomous vehicles (AVs) have the potential to revolutionize transportation, logistics, and environmental sustainability \cite{makahleh2024assessing} \cite{cortina2023fostering}. For instance, autonomous cars may decrease congestion and emissions  while guaranteeing safety~\cite{faisal2019understanding, narayanan2020shared} by improving coordination management~\cite{patire2024reduce}. Almost all studies on autonomous vehicles, including those referenced earlier, anticipates large-scale fleets of autonomous vehicles that operate in coordination to provide efficient transportation and logistics services.

In systems involving autonomous agents, where flows of agents intersect, traffic intersections are among the most critical for high performance~\cite{8732975}. Due to frequent agent interactions, efficient coordination management is considerably beneficial at these intersections.  Managing intersections in dense urban environments, particularly for autonomous agents \cite{lazarus2024multi} such as AVs, poses unique challenges. Urban space and road networks are constrained~\cite{doole2021constrained} by existing infrastructure~\cite{vascik2019assessing} and high traffic density\cite{hoekstra2001free,lowry2018towards,mueller2017enabling}, making the efficient coordination of agent movement critical for safety and operational efficiency.

Intersection management\cite{dresner2008multiagent,8732975,wu2022autonomous, nagrare2024intersection} has been extensively studied in both ground vehicle and airspace management domains, with the primary goals of minimizing congestion, reducing delays, and ensuring safety \cite{doole2020estimation}.
Approaches to intersection management are often classified into query-based and assignment-based systems, as noted in \cite{10.1145/3407903}. Centralized control offers a structured method where a single authority manages agent separation and trajectories but may struggle with diverse traffic patterns \cite{sunil2017analysis}. Hybrid or decentralized systems offer more flexibility \cite{sedov2018centralized}, though they may face coordination issues during high traffic or unforeseen circumstances. Research such as \cite{6859163} and \cite{del2007complexity} discusses managing controlled vehicles to avoid collisions, with scheduling approaches playing a pivotal role in ensuring safety \cite{10.1145/2883817.2883830,6760492}. {Nevertheless, the different sizes of the centralized control regions and their corresponding efficiency are not adequately addressed or studied.}

Signal-less intersection management of agents under centralized control has been explored  in\cite{wang2020cooperative,9133379}, where automated and connected vehicles\cite{DBLP:journals/corr/abs-2402-17043} coordinate themselves in the multi-intersection network to achieve conflict-free and efficient flow.{ Centralized intersection control systems are designed to maximize the number of agents passing through the intersection or minimize delay, reduce fuel consumption, and optimize the difference between expected and set arrival time\cite{9217470} \cite{9535372}. Additionally, Multi-level algorithm of intersection management \cite{li2023intersection} \cite{6094668} or control horizon for different crossing order \cite{10611649}} has been proposed. However, the importance of the optimal distance from the intersection, which is needed to control the agents efficiently while meeting all operational constraints, has not yet been investigated.

This paper investigates the effect of intervention timing in management strategies on maintaining operational constraints at intersections. We develop a framework for managing autonomous agents in constrained spaces, introducing a strategy called the "early versus late management" approach. In this context, "early" management involves proactive control of agents before they reach critical areas using larger control regions, while "late" management refers to intervention closer to critical areas with smaller control regions. This research aims to determine the optimal approach for maintaining operational constraints at intersections, primarily focusing on a better cross-flow management.

The main contribution of this work is a comparative analysis of early versus late management strategies at intersections, focusing on how these strategies improve the system's efficiency. By introducing the concept of control regions, represented as circles around the intersection, this study illustrates how intervention timing affect agents' coordination. Using a mixed-integer linear programming (MILP) model and receding horizon control to optimize the system's performance enables the analysis of natural platoon formation under these strategies, providing insights into minimizing delays while respecting operational constraints such as safety and intersection separation distances. The results emphasize the critical balance between early and late interventions, highlighting the significance of optimal control region sizing and timely management to ensure safe and efficient intersection operations, particularly in high-density traffic scenarios. While primarily focused on urban transportation, this paper's findings are relevant to other mobile agent systems, such as autonomous vehicles and drones. It aims to create a strategy to ensure safe and efficient operations by considering control region designs.

The rest of the paper is structured as follows. Section~\ref{sec_description} describes the problem setup and outlines the hypothesis, focusing on multi-agent behavior at an intersection and the influence of control region size on system performance. Section~\ref{model_formulation} details the MILP-based centralized control system. Section~\ref{sec_simulation} discusses simulation implementation with its key parameters like control region radii, velocity, and operational constraints. It also presents the results, focusing on performance metrics such as delay, runtime, and the impact of control region size on system efficiency. Finally, Section~\ref{sec_conclusion} concludes the paper and proposes directions for future research.

\section{Problem Description}\label{sec_description}

\begin{figure}[!ht]
    \centering
    \includegraphics[width=\columnwidth]{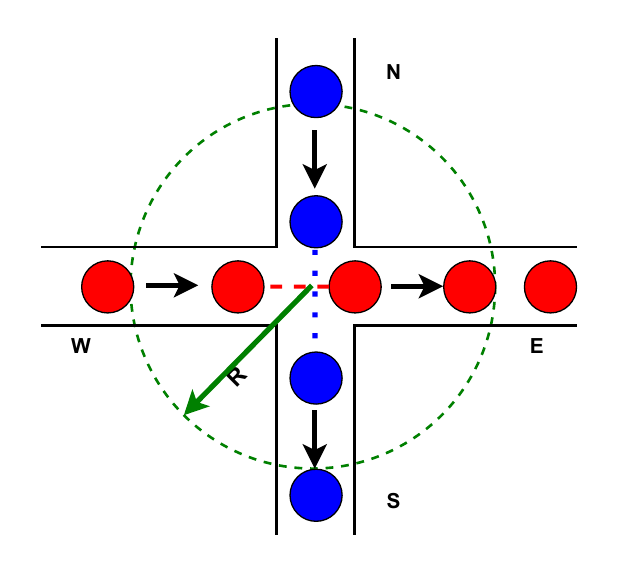}
    \caption{Illustration of control region circle of radius R for an intersection scenario.}
    \label{fig:frame_780_1}
\end{figure}

A model is created to analyze multi-agent behavior, navigating through a constrained space, specifically at an intersection. Each agent, a point mass whose dynamic is defined as the double integrator, travels along two orthogonal pathways and with the intersection at the origin. Agents move from north to south and from west to east, as shown in Fig.~\ref{fig:frame_780_1}. The influence of a centralized control system is represented by the control region radius \({R}\), which governs the management of agents. {The goal is to allow agents more autonomy until they are near the constraint, possibly improving efficiency, reducing unnecessary interventions, and efficiently using computational resources.}

\begin{Hypothesis}\label{hypo}Increasing the control region radius \({R}\) beyond a specific threshold results in diminishing returns on system performance.
\end{Hypothesis}
This hypothesis suggests that while enlarging the control region can initially enhance the efficiency and safety of agent management within constrained spaces; there is a critical limit. Beyond this limit, a further increase in the control region size fails to improve system performance proportionally. It emphasizes the importance of determining an optimal control region size to maximize system performance and safety.

\section{Model Formulation}\label{model_formulation}

The centralized control system is a {Mixed-Integer Linear Programming (MILP)} formulation \cite{schouwenaars2001mixed} \cite{KochAchterbergAndersenetal.2010} that maximizes the total velocity \({J_{v}}\)  of all agents across both pathways for total duration of the simulation. Simultaneously, it minimizes the total weighted deviation \({J_{p}}\) of a system's performance from its desired state across all agents and time steps. Mixed-integer linear programming is preferred due to the simplicity of implementation, and past successes in trajectory generation \cite{schouwenaars2005implementation}\cite{saber2023optimized} and mobile agents optimal space partitions\cite{mao2007space}. 

This centralized control system employs a receding horizon or model predictive control strategy\cite{garcia1989model}\cite{schouwenaars2004receding}, where the path of the agents is determined by sequentially solving the MILP for forthcoming time steps, creating locally optimal segments. However, only a subset of these computed commands are implemented per cycle. Typically, only the first set of control inputs is executed, akin to a classical approach, with a fresh set of commands recalculated at each time step\cite{schouwenaars2001mixed}.

\subsection*{Decision Variables} 
\begin{itemize}
    \item $s_{p,i}$: State variable of agent $p$ at time step $i$ .
    \item $u_{p,i}$: Control input of agent $p$ at time step $i$ .
    \item $w_{p,i}$: Slack variable for state deviation of agent $p$ at time step $i$ .
    \item {$V_{p,i}$:} Velocity state variable of agent $p$ at time step $i$.  
    \item ${A}_p$ and ${B}_p$ are the state transition and control input matrices
\end{itemize}

\subsection*{{Objective Function}}
\[
\left\{
\begin{aligned}
\text{Minimize} \;\; J_{p} &= \sum_{p=1}^P \left( \sum_{i=1}^{N-1} qo_p^\top \cdot w_{p,i} +  po_p^\top \cdot w_{p,N-1} \right) \\
\text{Maximize} \;\; J_{v} &= \sum_{p=1}^P \sum_{i=0}^{N-1} V_{p,i}
\end{aligned}
\right.
\]

\subsection*{Subject to}

\textbf{{1. State constraints:}}

$\forall p \in [1, P], \; \forall i \in [1, N]:$
\begin{align*}
s_{p,i} - s_{p,f} & \leq w_{p,i} \\
-s_{p,i} + s_{p,f} & \leq w_{p,i} \\
\end{align*}

\textbf{{2. Agent dynamics:}}

$\forall p \in [1, P], \;  \forall i \in [0, N-1]:$
\begin{align*}
    s_{p,i+1} &= A_{p} \cdot s_{p,i} + B_{p} \cdot u_{p,i} \\
    s_{p,0} &= (x_{p,0}, y_{p,0},  \dot{x}_{p,0}, \dot{y}_{p,0})^\top
\end{align*}

\textbf{{3. Control input limits:}}

$\forall p \in [1, P], \;  \forall i \in [0, N-1]:$
\begin{align*}
    u_{p,i} &\geq u_{p,\text{min}} \\
    u_{p,i} &\leq u_{p,\text{max}}
\end{align*}

\textbf{{4. State variable limits:}}

$\forall p \in [1, P], \;  \forall i \in [1, N]:$
\begin{align*}
    s_{p,i} &\geq s_{p,\text{min}} \\
    s_{p,i} &\leq s_{p,\text{max}}
\end{align*}

Where \({i}\) is the time index over  \({N}\) total time steps, each having \({\Delta t}\)   duration, and \({P}\) is the total number of agents. \({qo}\) and \({po}\) are non-negative weighting vectors for agent \({p}\).\\ \({s_{p,\mathbf{min}}}\), \({s_{p,\mathbf{max}}}\), \({u_{p,\mathbf{min}}}\) and \({u_{p,\mathbf{max}}}\) denote the minimum and maximum: state and control input vectors.

Additionally, the following constraints are considered to guarantee safe operations.

\textbf{{5. Safe distance constraint:}}
It ensures that all agents maintain a minimum safety margin from each other at all time steps. For every pair of agents \({p}\) and \({q}\) moving in the same direction, the constraint enforces a minimum distance of safe \({d_{safe}}\) between their positions at every time step $i$ throughout their navigation in the control region. This distance is maintained such that each agent $p$ must be at least safe \({d_{safe}}\) meters away from agent $q$, whether ahead or behind on the same trajectory. This ensures an agent respects the safety margin with the agent behind and ahead of it. Such a constraint is expressed as follows:\\

$\forall p \in [1, P], \;  \forall q \in [p+1, P], \;  \forall i \in [1, N]:$
\begin{align*}
    \quad s_{p,i,x} - s_{q,i,x} &\geq d_{\text{safe}}, \\
     \quad s_{q,i,x} - s_{p,i,x} &\geq d_{\text{safe}}, \\
     \quad s_{p,i,y} - s_{q,i,y} &\geq d_{\text{safe}}, \\
     \quad s_{q,i,y} - s_{p,i,y} &\geq d_{\text{safe}}, \\
\end{align*}
\textbf{{6. Intersection separation distance constraint:}}
It ensures safe navigation through the intersection for agents traveling along orthogonal pathways, meaning that collisions never occur by maintaining a separation distance at all times. For example, agent \({p}\) traveling eastbound and \({q}\) traveling southbound, computes their absolute distance from the intersection at each time step \({i}\), by calculating the norm-1 distance from their current positions to the intersection. Once these distances are determined, the model evaluates whether the sum of the distances of agent \({p}\) and \({q}\) meets or exceeds a specified minimum distance \({s_{dist}}\). This ensures that agents maintain a safe operational distance from agents traveling on the intersecting pathways. This constraint is only active as agents approach and traverse the intersection. Once agents have crossed the intersection, the agents no longer need to maintain this strict separation, as the risk of cross-traffic collisions diminishes.\\

$\forall p \in [1, P], \;  \forall q \in [1, P], \;  \forall i \in [1, N]:$
\begin{align*}
\intertext{where \( p \) is an agent traveling eastbound and \( q \) is an agent traveling southbound. \((x_c, y_c)\) are the coordinates of the intersection}
&d_{p,i} = |s_{p,i} - x_{\text{c}}| \\
&d_{q,i} = |s_{q,i} - y_{\text{c}}|\\
&|d_{p,i}| + |d_{q,i}| \geq s_{\text{dist}}
\end{align*}

The complete optimization formulation summarized in this sections \ref{model_formulation} is efficiently solved using the Gurobi Optimizer which is a software package \cite{gurobi} in Python.

\section{Simulation Framework}\label{sec_simulation}

We consider two flows: one originates from the north at +200m and stops at -200m in the South, and the other originates from the West at -200m and stops at +200m in the east. Agents are generated at random intervals with arrival times determined by the modulo operation between the current iteration and a random integer (1 to 7) multiplied by a random step size ${L}$, ensuring a non-uniform arrival distribution over ${T}/{\Delta t}$. To ensure the reproducibility of the random sequence, seed value was initialized (seed =20). The agents travel with maximum velocity ${V_{max}}$, and their positions are updated by a kinematic equation with constant ${V_{max}}$ when they are outside the circle. %

When the first agent enters the control region from any flow, the model starts running by adding the agent to the model, creating their variables, setting up their constraints, defining the objective function in the model, and running for specific horizon steps, as discussed in section \ref{model_formulation}. Once an optimized solution is found, each agent position is updated with their respective model solution. As more agents enter the control region from any flow, those are added to the model, create their variable, set constraints and objectives, and find the solution. The exact process is repeated for new agents that are added to the model. Agents are removed from the model or no longer considered by the model once they exit the control region circle. The complete trajectory of each agent is stored in a JSON file, which is used for post-processing and result generation. 

The Pseudocode \ref{alg:trajectory_simulation} describes the general functioning of the simulator, and the table [\ref{table:simulation_settings}] describes its initialization and parameter setting.

\begin{algorithm}
\caption{Agent trajectory simulation}\label{alg:trajectory_simulation}
\begin{algorithmic}
    \State Set ${\Delta t}$, ${L}, {R}, {N}$, ${V_{max}}$, ${d_{\text{safe}}}$, ${s_{\text{dist}}}$.
    
    \For{$i = 0, 1, \dots, \frac{{T}}{{\Delta t}}$}
        \State Add $p_{i}$ with $V_{max}$
        \If {position($p_{i}$) == R}
            \State Run MILP models:
            \State \hspace{1em} Add the objective and constraints[1-6]
            \State \hspace{1em} Update position($p_{i}$) from the model
        \EndIf
        \If {position($p_{i}$) == 2*R}
            \State Exit $p_{i}$ from MILP model
            \State Update position($p_{i}$)
        \EndIf
    \EndFor
\end{algorithmic}
\end{algorithm}

\begin{table}[!ht]
    \centering
    \caption{Simulation parameters and settings}
    \begin{tabular}{|p{4cm}|p{4cm}|}
        \hline
        {Parameter} & {Value} \\ \hline
        ${V_{max}}$: Maximum velocity & 15 [m/s] \\ \hline
        ${d_{\text{safe}}}$: Safe distance  & 3 [m] \\ \hline
       ${s_{\text{dist}}}$: Intersection separation distance & 4 [m] - 7 [m] \\ \hline

        ${\Delta t}$: Discretization step& 0.1 [s] \\ \hline

        {$L$}: Random step  & 3 \\ \hline
        
        $N$: Horizon steps  & 60 \\ \hline
        ${R}$: Radius   & [25, 40, 60, 90, 120, 150, 180] [m]\\ \hline
         ${T}$: Simulation duration & 150 [s] \\ \hline
        Start and End positions (N-S) & +200 [m] to -200 [m] \\ \hline
        Start and End positions (W-E) & -200 [m] to +200 [m] \\ \hline
        CPU Model & Apple M1 Max \\ \hline
        Solver (Python API) & Gurobi Optimizer \\ \hline
    \end{tabular}
    \label{table:simulation_settings}
\end{table}

\subsection{Simulation Results:}
Consider autonomous agents that travel on the orthogonal lanes and cross the intersection without collisions, as shown in Fig.~\ref{fig:frame_780_1}. Each agent is modeled as a discrete double integrator with a discretization step of 0.1s, with the constraints as  ${d_{\text{safe}}}=$ 3m and  ${s_{\text{dist}}}=$ 4m. For the results shown in this paper, the simulator was run for different sizes of control region radii, with fixed ${d_{\text{safe}}}$ value and varying ${s_{\text{dist}}}$ value as shown in table [\ref{table:simulation_settings}].  

\begin{figure}[!ht]
    \centering
     \begin{overpic}[scale=1]{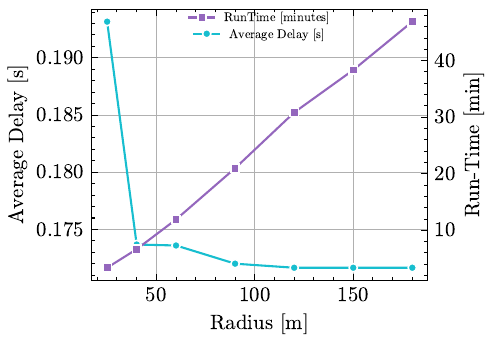}
           \end{overpic} 
    \caption{Average delay and runtime for different sizes of centralized control region for $302$ agents.}
    \label{fig:performanceof}
\end{figure}

Delay is one of the metrics that can be used to evaluate the performance of the different control region sizes. It is expressed as the difference between the actual time and the ideal time taken to cross the total diameter of the control region circle. Moreover, the ideal time is calculated as when the agent travels with maximum velocity to cover the same distance. Fig.~\ref{fig:performanceof} represents the average delay of different sizes of control region radius for the traffic pattern of 302 agents, which is the densest traffic the simulator can generate and handle. Fig.~\ref{fig:performanceof} demonstrates that as the radius size increases, the average delay gradually decreases until it reaches the optimal control region radius of around 120. The percentage of reduction of the delay from radius 25 compared to radius 120 is up to $12.52\%$. Beyond this optimal radius, the average delay remains quasi-constant.  

We observe that there is a relationship between the different sizes of the control regions and resource utilization, which can be measured using runtime, i.e., how long the centralized control system resources were running. It is shown in Fig.~\ref{fig:performanceof} that an increase in control region radius also increases the runtime, shown with purple square. Decreasing the delay by increasing the radius goes at the expense of the runtime. Hence, the optimal selection of the control region radius can be achieved by finding a good tradeoff between improving performance by increasing the radius and resource utilization.  

\begin{figure}[!ht]
    \centering
    \begin{overpic}[scale=1]{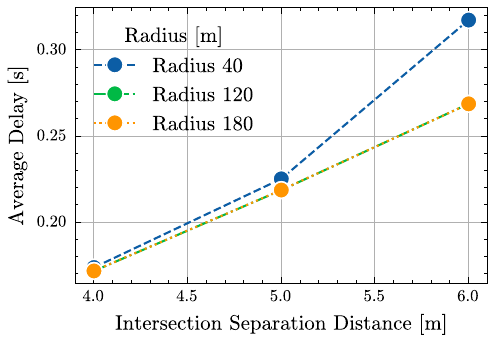}
           \end{overpic} 
    \caption{Average delay for different sizes of control region circles under the increasing intersection separation distance constraint. }
    \label{fig:Average_delay_size_intersection}
\end{figure}

Fig.~\ref{fig:Average_delay_size_intersection} represents the average delay for different control region radii of 40, 120, and 180 for increasing intersection separation distance ${s_{\text{dist}}}$ from 4 to 6. The average delay increases with the increase of ${s_{\text{dist}}}$ for all radii. For all the values of ${s_{\text{dist}}}$, there is a reduction of the delay from radius 40 to 120. However, a further increase in radius to 180 does not provide any better improvement. Hence, an optimal radius exists beyond which there is no decrease in average delay. This supports the hypothesis~\ref{hypo} stated above.

\begin{figure*}[h!]
    \centering
    \begin{overpic}[scale=0.34]{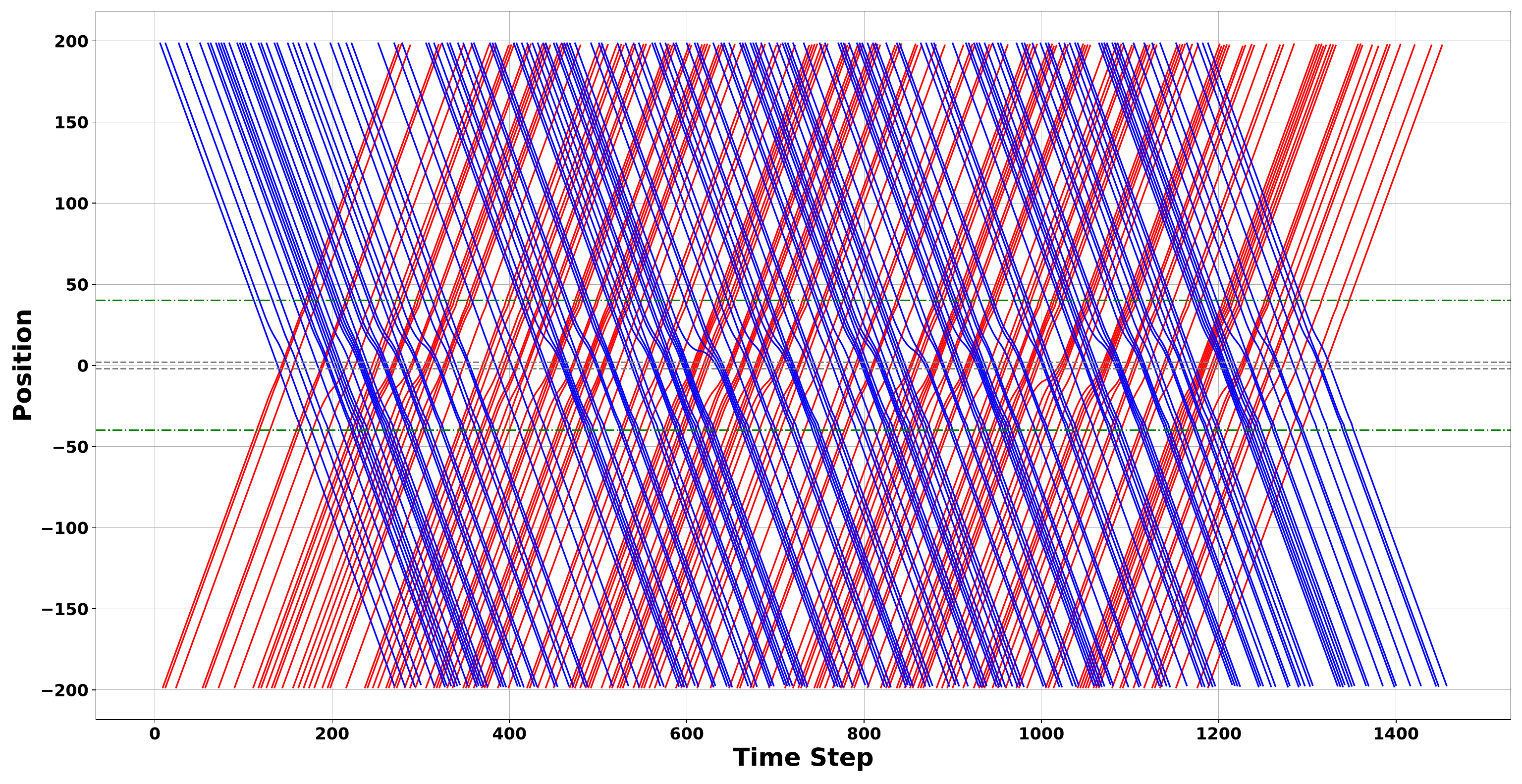}
    \end{overpic} 
    \caption{Space-time diagram for control region radius of 40m with (${d_{\text{safe}}}$) of 3m and  (${s_{\text{dist}}}$) of 7m }
    \label{fig:Space-Time40}
\end{figure*}

\begin{figure*}[h!]
    \centering
    \begin{overpic}[scale=0.34]{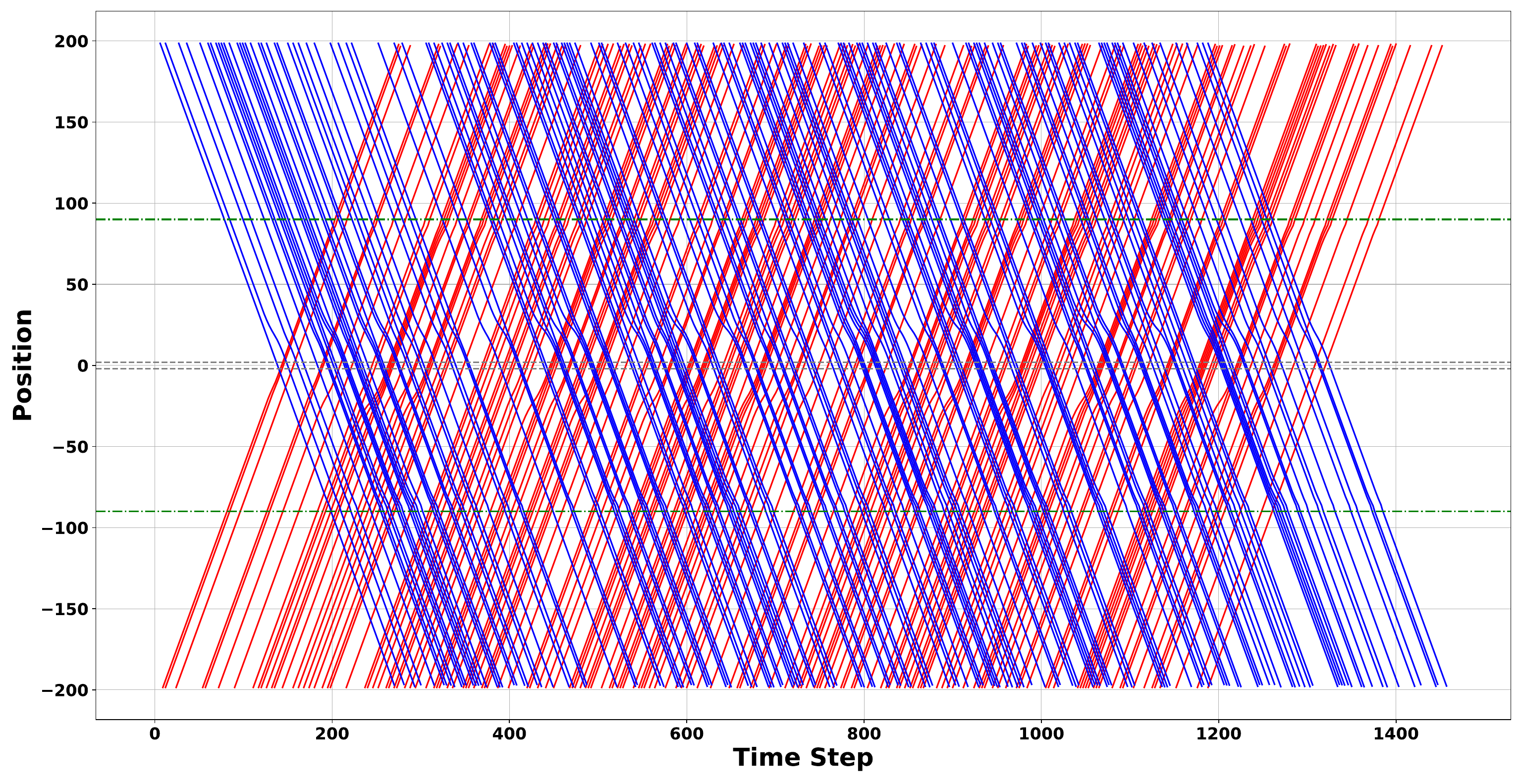}
    \end{overpic}
    \caption{Space-time diagram for control region radius of 90m with (${d_{\text{safe}}}$) of 3m and  (${s_{\text{dist}}}$) of 7m}
    \label{fig:Space-Time90}
\end{figure*}

The simulation model helps agents cross the intersection and avoid cross collision. It is like reinventing the traffic light by observing agents' trajectory near the intersection, where agents are slowing down and forming platoons to cross the intersection as shown in space-time Fig.~\ref{fig:Space-Time40} and Fig.~\ref{fig:Space-Time90}. Both figures show the complete journey of agents from start to end, which means from +200 to -200 meters from North to Southbound, and from -200 to +200 meters from west to eastbound. The green dashed line represents the radius, and the gray dashed line represents the intersection. We set which was set intersection separation distance to ${s_{\text{dist}}}=7$m. 

As shown in Fig.~\ref{fig:Space-Time40}and Fig.~\ref{fig:Space-Time90}, the agents are significantly slowing down to form a platoon. In radius $90$, there are tighter platoon formations, whereas radius 40 does not provide tighter platoons and is not optimal. The formation of the platoon happens due to the increase in intersection separation distance. Consequently, the delay will be reduced. When the control region size is large enough, then the deceleration of agents is minimal to form the optimal size of the platoon, which individually decreases the average delay as in Fig.~\ref{fig:Space-Time90}. 

{The GIF file of the simulation is available in the \href{https://github.com/salmanghori/Agent_control_region_circle/blob/main/README.md}{GitHub Repository README}. The GIF demonstrates the working of the simulator, while the space and time diagram related to the GIF is shown in Fig.~\ref{fig:Space-Time90}.}

\section{Conclusion}\label{sec_conclusion}

In this paper, we studied the effect of intervention timing in management strategies on maintaining operational constraints at intersections while guaranteeing safe separation distance and avoiding collisions. We introduced a strategy called the "early versus late management" approach. It consists of control regions represented as circles around the intersection. We used a mixed-integer linear programming model to optimize the system's performance. Our findings emphasize the existence of an optimal control region circle that significantly minimizes the delay at the intersection, beyond which there is no effective improvement in the flow. This understanding is paramount for optimizing agent management to improve system efficiency. Also, the selection of optimal control region at the expense of resource utilization, where a good tradeoff must be balanced per the requirement. Furthermore, the increase in separation distance value results in platoons' formation. The optimal and tighter platoon formations are possible in large control region sizes when compared to smaller size control regions, which leads to low delay value. The preliminary results provide initial evidence for the stated hypothesis in this paper.
 
Future work will focus on exploring additional metrics to further validate the hypothesis and identify key parameters influencing the optimal control region size. Additionally, the impact of early versus late management on the capacity of constrained spaces will be examined. An analytical model will also be developed to complement and enhance the simulation results.

\section*{Acknowledgment}
This research was funded by King Abdullah University of Science and Technology’s baseline support (BAS$/1/1682-01-01$).

\bibliography{name}
\end{document}